\documentclass[sigconf,nonacm]{acmart}

\usepackage{booktabs} 

\setcopyright{none}

\usepackage[inference]{semantic}
\usepackage{mathpartir}
\usepackage{mathtools}

\usepackage{bm}
\usepackage{listings}
\usepackage{todonotes}
\usepackage[noend]{algpseudocode}
\usepackage{algorithm}
\usepackage{multirow}
\usepackage{paralist}
\usepackage[utf8]{inputenc}
\usepackage{color}
\usepackage{caption}
\usepackage{empheq}
\usepackage{enumerate}
\usepackage{enumitem}

\usepackage{xspace}

\newcommand{\ant}[2]{$#1\xrightarrow{}{#2}$}
\newcommand{\antc}[3]{$#1\xrightarrow{#3}{#2}$}

\newenvironment{stable}[1]
    {\begin{table}[t]
    \begin{scriptsize}
    #1}
    {\end{scriptsize}
    \end{table}}

\newcommand{\rtit}[1]{\textsc{#1}}
  
\usepackage{supertabular}
\usepackage{ifthen}
\usepackage{alltt}
\usepackage{color}

\newenvironment{ottdefnblock}[3][]{ \framebox{#2} {~~~~~#3}}{}    

\newcommand{\ottnt}[1]{\mathit{#1}}

\newcommand{\ottkw}[1]{\mathbf{#1}}

\usepackage{latexsym}

\renewcommand{\mathbf}{\textbf}

\newcommand{\oolong}{OOlong\xspace}

\newcommand{\func}[1]{\textbf{\textsf{#1}\xspace}}

\newcommand{\off}[1]{}

\newcommand{\Rel}{\ensuremath{\mathit{Rel}}}
\newcommand{\Op}{\ensuremath{\mathit{Op}}}

\newcommand{\Id}{\ensuremath{\mathit{Id}}}
\newcommand{\Ids}{\ensuremath{\mathit{Ids}}}
\newcommand{\Cd}{\ensuremath{\mathit{Cd}}}
\newcommand{\Cds}{\ensuremath{\mathit{Cds}}}
\newcommand{\Msig}{\ensuremath{\mathit{Msig}}}
\newcommand{\Msigs}{\ensuremath{\mathit{Msigs}}}
\newcommand{\Fd}{\ensuremath{\mathit{Fd}}}
\newcommand{\Md}{\ensuremath{\mathit{Md}}}
\newcommand{\Fds}{\ensuremath{\mathit{Fds}}}
\newcommand{\Mds}{\ensuremath{\mathit{Mds}}}
\newcommand{\SB}{\{}
\newcommand{\FB}{\}}

\newcommand{\locate}{\ensuremath{\mathit{locate}}}
\newcommand{\Cfg}{\ensuremath{\mathit{cfg}}}
\newcommand{\eCfg}{\ensuremath{\mathit{ecfg}}}

\newcommand{\hcase}{\texttt{h}}

\newcommand{\eff}{\ensuremath{\mathit{eff}}}
\newcommand{\eqs}{\ensuremath{\mathit{eqs}}}
\newcommand{\sff}{\ensuremath{\mathit{sf}}}

\newcommand{\obj}{\ensuremath{\mathit{obj}}}

\newcommand{\mc}{\ensuremath{\mathit{mc}}}

\newcommand{\ret}{\triangleright}
\renewcommand{\c}[1]{\texttt{#1}}
\newcommand{\kw}[1]{\textbf{#1}\xspace}

\newcommand{\intk}{\kw{int}}

\newcommand{\weakk}{\kw{weak}}

\newcommand{\keyword}[1]{\textbf{#1}\xspace}

\newcommand{\thisk}{\keyword{this}}
\newcommand{\otherk}{\keyword{other}}

\newcommand{\nullk}{\keyword{null}}

\newcommand{\truek}{\keyword{true}}

\newcommand{\dom}{\func{dom}}

\newcommand{\snd}{{\sf snd}}

\newcommand{\model}{\textsc{Ant}\xspace}

\newcommand{\nope}[1]{}

\definecolor{mygreen}{rgb}{0,0.6,0}
\definecolor{mygray}{rgb}{0.5,0.5,0.5}
\definecolor{mymauve}{rgb}{0.58,0,0.82}
\definecolor{llightgray}{rgb}{0.93, 0.93, 0.93}

\makeatletter
\newcommand{\srcsize}{\@setfontsize{\srcsize}{6pt}{6pt}}
\makeatother

\lstdefinestyle{antoo}{
language=java,
commentstyle=\it,%
basicstyle=\srcsize\ttfamily,%
escapeinside={/*@}{@*/},
numbers=none,
numberblanklines=false,
extendedchars=true
breaklines=true,
showstringspaces=false,
mathescape,
frame=tb,
morekeywords={def, int, this, other, replicated, Unit, weak, implements},
backgroundcolor=\color{white},   
keywordstyle=\bfseries,
}

\newcommand\ocamlstyle{\lstset{%
language=[Objective]Caml,%
flexiblecolumns=false,%
keywordstyle=\color{blue},%
commentstyle=\it,%
basicstyle=\tiny\ttfamily,%
escapeinside={/*@}{@*/},
numbers=left,
numberstyle=\tiny,%
stepnumber=1,%
numbersep=15pt,%
extendedchars=true,%
breaklines=true,
showstringspaces=false,
mathescape=true,
frame=tb,
morekeywords={replicated, uptodate, repl, utd},
}
}

\lstnewenvironment{ocaml}[1][]{
\ocamlstyle
	\lstset{#1}
}
{}

\newcommand\ocamlexternal[2][]{{
	\ocamlstyle
	\lstinputlisting[#1]{#2}}
}

\newcommand{\lts}[1]
{ \setbox0=\hbox{$\ {\scriptstyle#1}\ $}
  \setbox1=\hbox{$\rightarrow$}
  \loop\setbox1=\hbox{$-$\kern-0.3em\unhbox1}\ifdim\wd1<\wd0\repeat
  \hbox{$\ \mathop{\box1}\limits^{#1}\ $}
}
\newcommand{\ltsr}[2]
{ \setbox0=\hbox{$\ {\scriptstyle#1}\ $}
  \setbox1=\hbox{$\rightarrow_{#2}$}
  \loop\setbox1=\hbox{$-$\kern-0.3em\unhbox1}\ifdim\wd1<\wd0\repeat
  \hbox{$\ \mathop{\box1}\limits^{#1}\ $}
}
\newcommand{\ltsrho}[1]
{ \ltsr{#1}\rho}

\newcommand{\ltsrs}[2]
{ \setbox0=\hbox{$\ {\scriptstyle#1}\ $}
  \setbox1=\hbox{$\rightarrow_{#2}^*$}
  \loop\setbox1=\hbox{$-$\kern-0.3em\unhbox1}\ifdim\wd1<\wd0\repeat
  \hbox{$\ \mathop{\box1}\limits^{#1}\ $}
}

\newcommand{\Lts}[1]
{ \setbox0=\hbox{$\ {}^{#1}\ $}
  \setbox1=\hbox{$\Rightarrow$}
  \loop\setbox1=\hbox{$=$\kern-0.3em\unhbox1}\ifdim\wd1<\wd0\repeat
  \hbox{$\ \mathop{\box1}\limits^{#1}\ $}
}

\newcommand{\nolts}[1]
{ \setbox0=\hbox{$\ {\scriptstyle#1}\ $}
  \setbox1=\hbox{$\centernot\rightarrow$}
  \loop\setbox1=\hbox{$-$\kern-0.3em\unhbox1}\ifdim\wd1<\wd0\repeat
  \hbox{$\ \mathop{\box1}\limits^{#1}\ $}
}

\begin{document}

\title{Anticipation of Method Execution in Mixed Consistency Systems -- Technical Report}

\renewcommand{\shorttitle}{}

\author{Marco Giunti,  Hervé Paulino and António Ravara}
\authornote{Work partially supported by the Portuguese Fundação para a Ciência e Tecnologia via project \textit{DeDuCe}~(PTDC/ CCI-COM/32166/2017) and NOVA LINCS (UIDB/04516/2020).
We thank Rúben Vaz for his contribution on the Java prototype implementation (\S~\ref{sec-java-impl}) used to apply \model to common use-cases.
This document is the companion technical report of paper~\cite{ant}.
}
\affiliation{%
  \institution{NOVA LINCS \& NOVA School of Science and Technology, NOVA University of Lisbon}
  \streetaddress{}
  \city{} 
  \state{Portugal} 
  \postcode{}
}
\email{marco.giunti@gmail.com,   {herve.paulino, aravara}@fct.unl.pt}

\renewcommand{\shortauthors}{}

\begin{abstract}
Distributed applications often deal with data with different consistency requirements: while a post in a social network only requires weak consistency, the user balance in turn has strong correctness requirements, demanding mutations to be synchronised. To deal  efficiently with sequences of operations on different replicas of the distributed application, it is useful to know which operations commute with others and thus, when can an operation not requiring synchronisation be anticipated wrt other requiring it, thus avoiding unnecessary waits.

Herein we present a language-based static analysis to extract at compile-time from code information on which operations can commute with which other operations and thus get information that can be used by the run-time support to decide on call anticipations of operations in replicas without compromising consistency. We illustrate the formal analysis on several paradigmatic examples and briefly present a proof-of-concept implementation in Java.
\end{abstract}

%
%



\maketitle

\section{Introduction}

Many Internet services replicate their data across multiple sites to improve availability,  fault-tolerance and scalability.
Data consistency in such systems is bound to the order on which operations are executed at each site, resulting in a tension between correctness and performance. 
Concerned about the latter, some datastores~\cite{cops-SOSP2011,dynamo-SOSP2007,cassandra-2010,riak,antidote} avoid global synchronization by resorting to weaker consistency semantics, allowing replicas to temporarily diverge~\cite{pec-Burckhardt2014}. 
There are however some operations (such as a withdraw from a bank account) whose correctness requires  consistency guarantees that cannot be specified in weakly consistent solutions.

\begin{figure*}
\includegraphics[width=\linewidth]{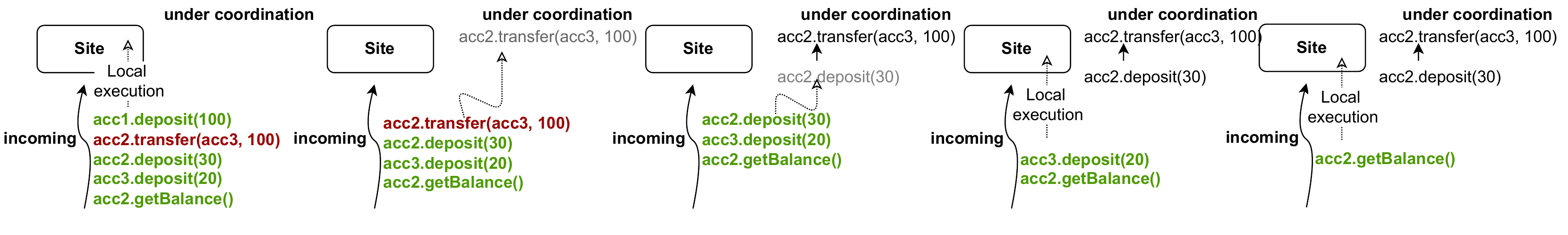}
\vspace{-10pt}
\caption{Bank accounts replicated among several sites. Calls that require coordination in red,  the remainder are marked green.}
\label{fig:example}
\end{figure*}

Consider  the   example of coordinating  operations over bank accounts  replicated at several sites, depicted in Fig.~\ref{fig:example}.%
\footnote{In the example the three account are: ``acc1'', ``acc2'' and ``acc3''.}
From the perspective of each site, given a sequence of incoming operations, it is fundamental to know which of these require
distributed coordination (in red) and which
 may be executed locally  first and subsequently propagated to the remainder sites (in green). 
The first operation, \texttt{acc1.deposit(100)}, does not require coordination and, hence, executes locally first.
Next, \texttt{acc2.transfer(acc3,100)}, in red, is placed under coordination to guarantee  its execution over the same state at every site.
It is followed by \texttt{acc2.deposit(30)}, which does not require coordination but cannot execute immediately because it may interfere with the transfer under coordination, 
causing the latter to execute over different states in different sites.
It is thus queued, waiting for the former to conclude.
This restriction does not, however, apply to \texttt{acc3.deposit(20)}, since it does not interfere with the execution of both  \texttt{acc2.deposit(30)} and \texttt{acc2.transfer(acc3,100)}.
So, it may execute locally first, ahead of operations received before and concurrently with the ones in execution (regardless of the coordination involved),
what we refer to as \emph{call anticipation}.
Finally, we have operation 
\texttt{getBalance()} that returns a weakly consistent view of an account's balance. Although operating over account \texttt{acc2}, the operation does not conflict with 
\texttt{acc2.deposit(30)} nor with \texttt{acc2.transfer(acc3,100)} and,\linebreak
hence, may execute locally first.

The state-of-the-art proposals on mixed consistency address the computation of the coordination level required by each operation executing over a distributed datastore, in order to guarantee the data's consistency requirements.
Works such as \cite{redblue-OSDI2012, se-POPL2016,fgc-ATC2018}  provided frameworks to identify which operations may execute without coordination.
%
Subsequently, others  have  focused on providing specifications that enable the automatic inference of the necessary coordination,  by 
requiring reasoning about  the side effects of each operation~\cite{ec-Eurosys2015}, and/or the visibility of data and  how all these interfere with each other~\cite{quelea-PLDI2015,fgc-ATC2018}.
%
%

Regarding the analysis of which operations may run concurrently, that gives rise to call anticipation,
the state-of-the-art~\cite{fgc-ATC2018,sieve-ATC2014,hamsaz-popl19} is to conduct a static commutativity analysis 
that produces a conflict table indicating if two operations may conflict, or not, instead of synthesizing the conditions 
when they conflict (or commute), unlike what is currently done in other contexts~\cite{DBLP:journals/jar/BansalKT20}.
Furthermore, the analyses work on semantic representations of the applications (either logical or flow-based models), not on actual code implementing the applications. They are language agnostic, but for evaluation purposes implement the approach in a particular setting.

In this paper we propose a language-based call anticipation static analysis named \model.
Rather than providing a generic semantic model, agnostic to the language in which the code is written (and sometimes even abstract on the way the model is built), we define our analysis directly on the code, with a language-based algorithm relying instead on formal definitions of program semantics.
This analysis is able to produce a conflict table that, for each pair of method calls, defines a system of inequalities that must be satisfied at runtime in order to anticipate a call.

We define our analysis over a augmented version of the \oolong~\cite{oolong-ACR2019}
core language, with constructs to
 (1) identify which fields are replicated, and of these, which allow for weak consistency,
 (2) the specification of invariants over the replicated fields, and
 (3) the definition of guards for method executions.
 The language shares similarities with MixT~\cite{mixt-PLDI2018} or DCCT~\cite{dccp-PLMDC2016} but distinguishes itself by not requiring the programmer to reason about consistency levels. Instead they must just classify fields as \textit{weak} or \textit{strong} (by default).
%

\emph{In short, our main contributions are the following}:
\begin{inparaenum}
%
\item 
A formal semantics  of the state conditions under which 
we can \emph{anticipate} a method call 
executed in a distributed runtime system featuring 
replicas~(\S~\ref{sec:conflicts});
\item A static analysis of object-oriented methods that builds on side effects and on their symbolic semantics to identify a system of inequalities over objects' fields and methods' parameters that must be satisfied at runtime in order to anticipate a call~(\S~\ref{sec:re-static-commutativity}).
\end{inparaenum}


\section{The Language}
\label{sec:model}
\begin{table}[t]
\centering
  \caption{Syntax of \model-\oolong programs.
    \textit{Ids, Cds, Fds, Mds and Msigs} are sequences of zero or
    more of their singular counterparts.
    Terms in grey background are only available at
  runtime.}
\label{tab:syntax}
\begin{tabular}{lcl}
  $P$ & $::=$ & \Ids~\Cds~$e$ \hfill \textit{(Programs)}
  \\
  \Id & $::=$ & \kw{interface} $I$ \SB \Msigs \FB 
      $~|~$ \\
      &&\kw{interface} $I$ \kw{extends} $I_1, I_2$ 
       \hfill \textit{(Interfaces)}
       \\
  \Cd & $::=$ & \kw{class} $C$ \kw{implements} $I$ \SB \Fds~\Mds \FB \hfill 
  \textit{(Classes)}
  \\
  \Msig & $::=$ & 
  $m(x : t_1) : t_2$ $[\tilde c]$ 
    \hfill \textit{(Signatures)}
  \\
  $d$ & $::=$ & $x$ $~|~$ $x.f$  $~|~$ $v$ 
  \hfill \textit{(Invariant values)}
  \\
  $c$ & $::=$ & 
  $d_1 \ \Rel\  d_2$ 
  \qquad\qquad\hfill \textit{(Invariants)}
  \\
  \Md & $::=$ & \Msig \SB$e$\FB
  \hfill \textit{(Methods)}\\
  \Fd & $::=$ & $f : t \sim\hspace{-.2em}\weakk~[\tilde c]$ \hfill \textit{(Fields)} \\
  $e$ & $::=$ & $sv$
        $~|~$ $x.f$
        $~|~$ $x.f$ \c{=} $e$
        $~|~$ $x.m(e)$ 
         $~|~$    \\
         &&\kw{let} $x$ \c{=} $e_1$ \kw{in} $e_2$
        $~|~$ \kw{new} $C$
        $~|~$ $(t)~e$\qquad\qquad
        \hfill \textit{(Expressions)} 
      \\
  $v$ & $::=$ & 
  \kw{null}
  $~|~$ $n\in\intk$
  $~|~$ \colorbox{lightgray}{$\iota$} \hfill \textit{(Values)} \\
  $sv$ & $::=$ & $v$
  $~|~$ 
  $x$
  $~|~$ $sv_1 ~ \ottnt{Op} ~ sv_2$
  \hfill \textit{(Symbolic Values)} \\
   $t$ & $::=$ & $I$ $~|~$ $C$ $~|~$ $\mathbf{Unit}$ $~|~$ $\intk$\hfill
  \textit{(Types)}
                   \\
                 $\Gamma$ & $::=$ & $\epsilon$
  $~|~$ $\Gamma, x : t$
  $~|~$ \colorbox{lightgray}{$\Gamma, \iota : C$} \hfill
                     \textit{(Typing Environment)}
                      \\
     \colorbox{lightgray}{$\Cfg$} & $::=$ & $\langle H, V, \sff, \tilde c, ST\rangle$\qquad\qquad\qquad \hfill \textit{ { (Configuration)}}
     \\
     \colorbox{lightgray}{$H$} & $::=$ & $\epsilon$ $~|~$ $H, \iota \mapsto \obj$\hfill \textit{ { (Heap)}}
     \\
     \colorbox{lightgray}{$V$} & $::=$ &$\epsilon$ $~|~$ $V, x \mapsto v$\hfill  {\textit{(Stack)}}
     \\
   \colorbox{lightgray}{$\sff$}& $::=$ & $\epsilon$ $~|~$ 
   $\sff,(\iota, f,\kw r)$ $~|~$
   $\sff,(\iota, f,\kw w)$ \hfill
                    {\textit{(Strong fields)}}
     \\
     \colorbox{lightgray}{$ST$} & $::=$ &$e$ $~|~$ $\kw{EXN}$\hfill  {\textit{(Single Threads)}}
     \\
     \colorbox{lightgray}{$\obj$} & $::=$ &$(C, F)$\hfill  {\textit{(Objects)}}
     \\
     \colorbox{lightgray}{$F$} &$::=$ &$\epsilon$ $~|~$ $F, f \mapsto v$\hfill  {\textit{(Field map)}}
\end{tabular}
\end{table}





The object-oriented language under consideration, 
dubbed \model-\oolong, is meant to model replicated systems where object fields are \textit{strongly consistent},
unless explicitly marked as weakly consistent.
We present in Table~\ref{tab:syntax} the syntax of
\model-\oolong.

\paragraph*{Syntax.}
So, a field declaration, by default strongly consistent, can be optionally flagged with the \weakk keyword
to indicate that the field has weak consistency. 
Field and method declarations are decorated with a (possibly empty) set of invariants~$c$, 
where in the latter case $c$ is often referred to as a precondition;
an invariant is an application of a binary predicate \Rel\ over integers.
Invariants are used solely for program analysis purposes 
(the call anticipation analysis of \S~~\ref{sec:conflicts} and
\S~\ref{sec:static-commutativity}),
and have no impact on the dynamic semantics. 

A value $v$ can be a \nullk~reference, an integer~$n$ or a location~$\iota$.
A novelty is the presence of \emph{symbolic values}~$sv$ that include values, variables $x$ and applications of integers operators $\Op$ to symbolic values.
As usual, occurrences of a variable $x$ are \emph{bound} in an expression $e_2$ in the context of a let-expression \kw{let} $x$ \c{=} $e_1$ \kw{in} $e_2$. Occurrences of other variables in $e_2$, 
or in general in an expression not inside a let are \emph{free}.


\paragraph*{Semantics.}
 \begin{table}
\caption{Closure of preconditions 
\framebox{$\func{closure}_{H,V}(\tilde c)=\tilde c$}}
\label{tab:closure}
\begin{mathpar}
\inference{
    V(x) =v
}{
    \func{closure}_{H,V}(x)=v    
}
\and
\inference{
    V(x) =\iota\qquad H(\iota)=(C,F)\qquad F(f)=v
}{
    \func{closure}_{H,V}(x.f)=v    
}
\and
\inference{}{
    \func{closure}_{H,V}(v)=v
}
\and
\inference{
    \func{closure}_{H,V}(d_1)=v_1  \qquad
    \func{closure}_{H,V}(d_2)=v_2  
}{
    \func{closure}_{H,V}(d_1\,\Rel\, d_2)=v_1\,  \Rel\, v_2
} 
\and
\inference{
    \func{closure}_{H,V}(c_1)=c_1'  \cdots
    \func{closure}_{H,V}(c_n)=c_n'}{
        \func{closure}_{H,V}(c_1,\dots,c_n)=c_1',\dots,c'_n
    }
\end{mathpar}    
\end{table}    
    \begin{figure*}
        \caption{Semantics of \model-\oolong programs 
        \framebox{$\Cfg \hookrightarrow \Cfg'$}} 
        \label{fig:dynamics}
      \begin{tabular}{lcl}
        $E[\bullet]$ & $::=$ & $x.f = \bullet$ $~|~$
        $x.m(\bullet)$ $~|~$
        $\bullet\ \ottnt{Op}\ sv$ $~|~$   
        $sv\ \ottnt{Op}\ \bullet$ $~|~$
        \kw{let} $x$ \c{=} $\bullet$ \kw{in} $e$
        $~|~$  $(t)\,\bullet$
      \end{tabular} 
      \\
    \begin{mathpar}
      \inference[\rtit{dynEvalCxt}]{
        \langle H, V, \sff, \tilde c, e  \rangle
        \hookrightarrow
        \langle H', V', \sff', \tilde c', e'\rangle
        }{
        \langle H, V, \sff, \tilde c, E[e]  \rangle
        \hookrightarrow
        \langle H', V', \sff', \tilde c', E[e']}
        \and
        \inference[\rtit{dynEvalOp}]{
        n = n_1 ~\ottnt{Op}~ n_2
        }{
        \langle H, V, \sff, \tilde c, n_1 ~\ottnt{Op}~n_2  \rangle
        \hookrightarrow
        \langle H, V, \sff, \tilde c, n \rangle
        }
        \and
        \inference[\rtit{dynEvalVar}]{
        V (x) = v
        }{
        \langle H, V, \sff, \tilde c, x\rangle \hookrightarrow
        \langle H, V, \sff, \tilde c, v\rangle
        }
    \and
        \and
        \inference[\rtit{dynEvalSelect}]{
        V (x) = \iota \qquad
        H (\iota) = (C, F)\\
        F(f)=v\qquad
        \sff' =\sff,(\iota, f, \kw r)
        }{
        \langle H, V, \sff, \tilde c, x.f\rangle \hookrightarrow
        \langle H', V, \sff', \tilde c, v\rangle
        }
    \and
    \inference[\rtit{dynEvalUpdate}]{
    V (x) = \iota \qquad
    H (\iota) = (C, F)\\
    H'= H[\iota\mapsto (C, F[f\mapsto v])]\qquad
    \sff' =\sff,(\iota, f,\kw w)
    }{
    \langle H, V, \sff, \tilde c, x.f = v\rangle \hookrightarrow
    \langle H', V, \sff', \tilde c, \nullk\rangle
    }
    \and
    \inference[\rtit{dynEvalNew}]{
    \kw{fields}(C)= f_1:t_1\sim\hspace{-.2em}\weakk~[\tilde c_1],\dots,
    f_n:t_n\sim\hspace{-.2em}\weakk~[\tilde c_n]
    \\
    F = f_1\mapsto \func{init}(t_1),\dots,f_n\mapsto\func{init}(t_n)
    \qquad\iota \textit{ fresh}}{
    \langle H, V, \sff, \tilde c, \kw{new}\ C\rangle 
    \hookrightarrow
    \langle H[\iota\mapsto(C,F)], V, \sff, \tilde c, \iota\rangle
    }
    \and
    \inference[\rtit{dynEvalLet}]{
    x' \text{ fresh}\qquad
    V' = V[x'\mapsto v] \qquad
    e' = e [x \mapsto x'] }{
    \langle H, V, \sff, \tilde c, \kw{let}\ x = v\ \kw{in}\ $e$\rangle 
    \hookrightarrow
    \langle H, V', \sff, \tilde c, e'\rangle
    }
    \and
    \inference[\rtit{dynEvalCast}]{
    }{
      \langle H, V, \sff, \tilde c, (t) v\rangle 
      \hookrightarrow  
      \langle H, V, \sff, \tilde c, v\rangle
    }  
    \and
    \inference[\rtit{dynEvalCall}]{
    V (x) = \iota \qquad
    H (\iota) = (C, F)\qquad
    m(y : s) : t' \,[\tilde c']\{ e \}\in \func{methods}(C)\\
    x\not\in{\bf bv}(e)\qquad y'\textit{ fresh}\\
    \\
    V' = V[y'\mapsto v]\qquad
    \qquad
    e' = e[\thisk \mapsto x][y\mapsto y'] 
    }{
    \langle H, V, \sff, \tilde c, x.m(v)\rangle \hookrightarrow
    \langle H, V', \sff, (\tilde c, \func{closure}_{H, [y\mapsto v]}(\tilde c')), e'\rangle
    }
    \end{mathpar}
    \end{figure*}

The evaluation relation, noted $\Cfg\hookrightarrow \Cfg'$, 
recast the single-thread semantics of OOlong, 
modulo the presence of integer operations.
The rules are in  defined in Table~\ref{tab:semantics}.
We omit the rules for exceptions, which are those defined in~\cite{oolong-ACR2019}.
A configuration \Cfg\ is pair $\langle \Sigma, e\rangle$ where $\Sigma$ is the {\it state}
and $e$ is an expression. 
A state contains a heap~$H$ mapping locations to objects, 
a stack $V$ mapping variables to values,  
and two extra-fields to {\it monitor} the method calls' preconditions 
and the access to strong fields, respectively.

The first extra-field monitors if an execution respected the 
method's preconditions:
e.g. if a method $m$ of class $C$ has the pre-condition $\thisk.f \geq 1$,
where $f$ is an integer field of $C$,  and $m$ is executed, 
then the first extra-field contains
the entry $n\geq 1$, where $n$ is the value of $\thisk.f$ at execution time;
by evaluating $n\geq 1$ \emph{after} the call's execution, 
we are able to infer if the call was legit.
The second extra-field contains entries of the form $(\iota, f, \kw r)$ and 
$(\iota, f, \kw w)$ indicating that
the strong field $f$ of the object in location $\iota$ has been accessed 
during the execution in \emph{read} and \emph{write} mode, respectively.


An object is a pair $(C,F)$, where $C$ is a class identifier and $F$ is a map from fields to values.

\paragraph*{Type System.}
The typing rules for expressions are those of~\cite{oolong-ACR2019} 
extended with the rules for integers. 
The rules inductively define the typing relation $\Gamma\vdash e\colon t$.
The typing rules for fields and methods enforce that 
an invariant $c$ of a field $f$ cannot refer to fields different from~$f$, and 
a precondition $c$ of a method cannot refer to weak fields, respectively.
The typing rules for configurations, noted as $\Gamma\vdash\Cfg\colon t$,
extend straightforwardly~\cite{oolong-ACR2019} by adding the rules for the monitoring fields.

All typing rules are available in~\S~\ref{sec:typing-rules}.

\begin{lstlisting}[caption=Bank account, label=lst:bank,float]
class Account implements Object {
  balance  : int weak [this.balance$\geq$0]
  min_cash : int [this.min_cash$\geq$50]
        
  def init (amount : int) : Unit [amount$\geq$this.min_cash]{ 
   this.balance = amount 
  }
         
  def withdraw(amount : int) : Unit [amount$>$0] { 
   this.balance -= amount  
  } 
        
  def transfer(recv : Transfer) : Unit {
   let a = recv.amount in let w = recv.account in
    this.withdraw(a); w.deposit(a) 
  }
        
  def deposit(amount : int) : Unit [amount$>$0] { 
   this.balance += amount
  } 
          
  def accrueInterest(interest : int) : Unit {
   let x = this.balance * interest / 100 in
    this.balance += x
  }
            
  def getBalance() : int  { this.balance }
  
}
\end{lstlisting}

\paragraph{Programming Example}
Listing~\ref{lst:bank} shows how to program  in \model-\oolong an account whose balance is replicated but does not require strong consistency.
Importantly, as intended, the code is identical to a non-replicated implementation.
The weak field \texttt{balance} cannot be mentioned in the methods' preconditions;
on contrast, the strong field \texttt{min\_cash} can be accessed in method \texttt{init}
to verify that the initial amount is greater or equal to the minimum allowed.
The state invariant is the invariant of fields \texttt{balance} and \texttt{min\_cash}.
Methods that do not change the state invariant are \textit{locally permissible} 
(cf.~\texttt{getBalance}), while a method~$m$ that may change the state invariant is locally permissible in a pre-state satisfying its invariants if and only if all calls of $m$ satisfying the preconditions spawned from the call lead to post-states that satisfy their invariants; for instance, a call $\texttt{transfer}(n)$ spawns the precondition $n>0$.
Therefore, methods \texttt{init} and \texttt{deposit} are locally 
permissible, since the amount to be initialized or deposited, respectively, 
must be positive, while \texttt{withdraw}, \texttt{transfer} and 
\texttt{accrueInterest} are not locally permissible.

\paragraph{Method Call Execution}
\model method calls execute as indivisible units over the  target data. \S~\ref{sec:cs} will elaborate on the consistency level assigned to the execution of each method call.int
For now, to make the presentation  of the following sections clearer, we  simply state that method calls that either access strong fields or are not locally permissible require strong consistency semantics.
The remainder may execute under causal or even eventual consistency, depending on whether causality relations have been identified between  methods or not. 

\section{Method Call Anticipation}\label{sec:conflicts}
This section introduces the runtime notion of \emph{method call anticipation}, 
which defines, given a sequence of method calls executed in a distributed runtime system featuring replicas,
the \emph{state} conditions under which one of the calls can be safely anticipated in the sequence, without leading to possible inconsistent state in different replicas executing that same sequence.
To this aim, taking inspiration from Hamsaz~\cite{hamsaz-popl19},
we recast the notion of local permissible methods and of commutative 
method calls in \model, and we equip them with a formal semantics with respect to our language.

\paragraph*{Key notions}
Before defining call anticipation, we present the main concepts we build on.
In the remainder, we use $\mc$ to range over expressions $x.m(sv)$, and
consider configurations $\langle \Sigma,\mc\rangle$, where we assume that
there exists $\Gamma$ and $t$ s.t. $\Gamma\vdash \langle \Sigma,\mc\rangle\colon t$.
Whenever there are a state $\Sigma'$ and a value $v$ s.t. 
$\langle \Sigma,\mc\rangle\hookrightarrow^*$
$\langle \Sigma', v\rangle$, we let
$\func{update}(\mc,\Sigma) =\Sigma'$.
We denote single substitutions with $\mapsto$, and, given an invariant value $d$ and a state $\Sigma$, 
let the substitution $d\Sigma$ be equal to $v$ when $d = x$ and $V(x)=v$, or $d= x.f$ and $V(x)=\iota$ and $H(\iota) = (C, F)$ and $F(f)=v$, and be equal to $d$ otherwise,
where $H$ and $V$ are the heap and the stack of $\Sigma$, respectively.

\begin{description}
  \item[Satisfied Invariant]
  An invariant $d_1 \Rel\ d_2$ is satisfied by a state $\Sigma$, noted $\Sigma \Vvdash d_1 \Rel\ d_2$, whenever $d_1\Sigma\ \Rel\ d_2\Sigma = \truek$.
  \item[State Invariant]\label{def:state-invariant}
  Let $H$ and $V$ be the heap and the stack of a state $\Sigma$, respectively. The invariant predicate $\textbf I(\Sigma)$ holds whenever
  $V(x)=\iota$ and $H(\iota)=(C,F)$ and 
  $f : t \sim\hspace{-.2em}\weakk~[\tilde c]$ is a field declaration of class $C$ and
  $c\in\tilde c$ imply
  $\Sigma \Vvdash c[\thisk\mapsto x]$.
  \item[Call Guard]
  Let $\Sigma$ be a state with empty monitored preconditions, and  
  assume  
  $\func{update}(\mc, \Sigma) =\Sigma'$.
  A call $\mc$ is $\Sigma$-guarded
  when $c$ is a monitored precondition of $\Sigma'$ implies
  $\Sigma'\Vvdash c$.
\end{description}

We are interested in analyzing the permissibility of a guarded call in a given pre-state,
that is when the post-state spawned by the call preserves the
invariant of the pre-state.   
Local permissibility of a method requires that all its guarded calls are permissible.

\begin{description}
  \item[Permissible and Not-Permissible Calls]
  Let  $\mc$ be a $\Sigma$-guarded method call.
  We say that $\mc$ is \emph{permissible} in $\Sigma$,
  written as $\textbf{P}(\Sigma, \mc)$,
  if $\textbf I(\Sigma)$ implies 
  $\textbf I(\func{update}(\mc, \Sigma))$; and we say that

  $\mc$ is \emph{not permissible} in $\Sigma$,
  written as $\textbf{NP}(\Sigma, \mc)$, 
  if $\textbf I(\Sigma)$ implies 
  $\neg \textbf I(\func{update}(\mc, \Sigma))$;
  \item[Locally Permissible]\label{def:lp}
  A method $m$ is locally permissible in a state $\Sigma$,  written as
  $\textbf{LP}(\Sigma, m)$, if and only if,
  
  all $\Sigma$-guarded calls of $m$ are permissible in $\Sigma$.
\end{description}

%

State-commutativity  and State-conflict are symmetric binary relations over
method calls indexed by states.
Rather than requiring,  as in~\cite{hamsaz-popl19},
that commuting method calls enforce this
behaviour in all possible states, we tailor our analysis and
definition to the state under consideration.

\begin{description}
  \item[State-Commutativity and State-Conflict]\label{def:s-conflict}
  Let $\Sigma$ be a state. Two method calls $\mc$ and $\mc'$ $\Sigma$-{commute} if and only if

  $
  \func{update}(mc', \func{update}(mc, \Sigma))
         = 
  \func{update}(mc, \func{update}(mc', \Sigma))
  $
  Otherwise, they $\Sigma$-{conflict}. 
\end{description}

We are now ready to introduce the (as far as we know, original) notion of \emph{call anticipation}, stating:

\emph{when can a sequence of two method calls $\mc;\mc'$ be reordered
into $\mc';\mc$, thus anticipating the call $\mc'$}.

To this aim, we rely on
\begin{inparaenum}
  \item quantification over integers instantiating the weak consistency fields, and
  \item the fields that must preserve strong consistency within a method call~$\mc$.
\end{inparaenum}
Quantification is needed since, in our analysis, we cannot trust the actual value of fields with weak consistency, because they could be outdated. 
The {\it weak integer fields' generalization} of a state $\Sigma$ w.r.t. a tuple of integers $N=n_1,\dots,n_m$,  noted
$\func{wifg}(\Sigma, N)$, 
is obtained by substituting the
values in the field map of the objects in the heap, such that the field is an integer tagged with
$\weakk$ with the integers in $N$, where we assume that the are $m$ such values.

\begin{table}
  \caption{Call Anticipation Algorithm}
  \vspace{-10pt}
  \label{tab:anticipation-algorithm}
  \ocamlexternal{alg.ml}
\end{table}

\paragraph*{The anticipation algorithm} is presented in Listing~\ref{tab:anticipation-algorithm}:
given a state $\Sigma$ and a sequence $\mc_1;\mc_2$, it says when
we are allowed to anticipate the call $\mc_2$ w.r.t. $\mc_1$.

The algorithm requires that the two calls are commutative (l2);
subsequently, we have the following steps:
\begin{enumerate}
\item check whether the call $\mc_2$ is locally permissible in the pre-state $\Sigma$ or if
the call is permissible for all possible values of the weak integer fields of $\Sigma$ (l4);
\item check whether for all possible values $n_1,...,n_m$ of the weak integer fields of $\Sigma$,
the call $\mc_1$ is locally permissible in the post-state, 
that is the update of $\func{wifg}(\Sigma, (n_1,...,n_m))$ with $\mc_2$ (l6--9);
\begin{enumerate}
  \item if not, check whether the permissibility of $\mc_1$ has not changed in the pre and in the post states:
  \begin{itemize}
    \item if $\mc_1$ was permissible in the pre-state, it must be permissible in the post-state (l11)
    \item if $\mc_1$ was not permissible in the pre-state, it must be non-permissible in the post-state (l12);
  \end{itemize}
\end{enumerate}
\item check whether  the value of 
    strong fields accessed in \emph{read} mode in the execution of 
    $\langle \func{update}(\mc_1,\Sigma), \mc_2\rangle$ 
    are not accessed in \emph{write} mode during the execution of $\langle \Sigma,\mc_1\rangle$ (l13).
\end{enumerate}

To see why the double implication is necessary in 2(a),
consider the case where there are two replicas and  a sequence of calls $\mc_1;\mc_2$,
and  where $mc_1$ is both non-permissibile in the first replica and
permissible in the second replica, because the two replicas are accessing weak fields with different values. 
The anticipation of $\mc_2$ should preserve this invariant:
that is, after the anticipation of $\mc_2$,
the call $\mc_1$ should still be non-permissibile in the first replica
and non-permissible in the second replica, 
otherwise the two replicas would diverge after performing the call anticipation.

We stress that both checks 1) and 2) must be effectuated irrespectively of the actual values of the weak 
(integer) fields in the heap, hence the usage of universal quantification.

\paragraph*{Limitations}
The main obstacle to use the algorithm at runtime is the overhead:
the use of quantification would require the help of a constraint solver, which in most cases 
would make unfeasible to obtain an answer in a reasonable time.
This motivates the need of the static analysis introduced in \S~\ref{sec:re-static-commutativity}:
it returns a system of inequalities over integer fields and method's parameters that can be
efficiently checked at runtime in order to decide anticipation. 

\begin{example}\label{ex:anticipation}
  Consider the methods defined in Listing~\ref{lst:bank}, the sequence 
  $x.\texttt{withdraw}(5);$ $x.\texttt{deposit}(10)$, and a state $\Sigma$ where the
  heap contains, for some integer $n'\geq 0$, the entry
  \begin{center}
  $x\mapsto(A,F[\texttt{balance}\mapsto n'][\texttt{min\_cash}\mapsto 50])$
  \end{center}
  \noindent
  We apply the algorithm steps to decide if we can obtain the sequence 
  $x.\texttt{deposit}(10);$ $x.\texttt{withdraw}(5)$. 
  The methods commute, as the operations $+$ and $-$ do.
  \texttt{Deposit} is LP, therefore step (1) is fulfilled.
  In step (2), $\func{wifg}(\Sigma, n)$ contains a heap with an entry
  $x\mapsto(A,F[\texttt{balance}\mapsto n][\texttt{min\_cash}\mapsto 50])$ (the weak value $n'$ has been 
  substituted with an arbitrary $n$).
  Thus, \texttt{withdraw} is not LP in this state, since a call
  $x.\texttt{withdraw}(a)$ with $n<a$ would not be permissible, as it would break
  the post-state invariant.

  The next step is the analysis of sub-case (2.a):
  we need to establish that if the call $x.\texttt{withdraw}(5)$ is 
  (not-)permissibile in the pre-state, then it is (not-)permissible in the post-state, respectively.
  This does not hold: e.g. when $n < 5$,
  we have that $x.\texttt{withdraw}(5)$ is not permissible, while it is permissible
  in the post-state, since $n\geq 0$ implies $5< n+10$, and in turn the post-state satisfies its invariant.
  The algorithm therefore returns false.
  \end{example}

\section{Static Semantics of  Anticipation}\label{sec:static-commutativity}
\label{sec:re-static-commutativity}
In this section, we present a source-level analysis of \model programs.
The aim 
is the \emph{automatic generation of the conditions} under which,
given a sequence of method calls $\mc_1;\mc_2$ executed in a distributed runtime system featuring replicas,
we can anticipate the call $\mc_2$ w.r.t. $\mc_1$, leading to the sequence
$\mc_2;\mc_1$.
The conditions are expressed in terms of a system of inequalities populated by aliases of the integer fields, and of the actual (integer) parameters.

The analysis has three levels -- we identify:
\begin{enumerate}
  \item the conditions under which pairs of methods can give rise to commutative calls;
  \item locally permissible methods that do not need any coordination at runtime;
  \item and, when a method is not LP, the conditions under which its call can be anticipated.
\end{enumerate}

\subsection*{Effect Inference}\label{sec:effects}

\begin{stable}
  \caption{Effects of typed methods}\label{tab:effects-inference}
     \textit{Syntax of effects}\qquad
\begin{tabular}{lcl}
    $x_b$ & ::= &  $\ottkw{VbindL}(x) ~|~
    \ottkw{VbindC} (x, \tilde c) ~|~ 
    \ottkw{Vfield}(x,f)$ 
                 \hfill \textit{(Effect Variables)}
    \\
    $b$ &::=& $\ottkw{Evar}(x_b) ~|~ \ottkw{Eop}({Op}, e)   ~|~
                   \ottkw{Eret}(e)  ~|~
                   \ottkw{Enew}(C,\Fds) 
$ \qquad\qquad\qquad \hfill \textit{(Effect Expressions)}
     \\
    $\textit{E}$ & $::=$ & $[]$ $~|~$ $b :: E$ \hfill \textit{(Effect Expression Lists)}
    \\
    $\eff$ & $::=$ & $(\tilde c, E)$ 
                     \hfill \textit{(Effect)}
  \end{tabular}
\\
    \textit{Static inference rules for expressions and methods.} 
    \qquad \framebox{$\Gamma\vdash e \ret E$}
    \framebox{$\Gamma\vdash_x \Md \ret \eff$}
    \tiny
  \begin{mathpar}
 
     \inference [\rtit{eOp}]{
    \Gamma\vdash sv_1\ret E_1
    }{
    \Gamma \vdash sv_1 \ottnt{Op}~ sv_2 \ret  
    E_1 \circ [\ottkw{Eop}(Op,sv_2)]
    }
    \and
  \inference[\rtit{eNew}]{
  }{
  \Gamma\vdash\kw{new} \ C \ret [\ottkw{Enew}(C,\kw{fields}(C))]
  }
  \and
  \inference[\rtit{eCall}]{
    \Gamma\vdash sv \ret E_1\quad
    \Gamma(x) = C \quad
   m(y : s) : t' \,[\tilde c]\{ e \}\in \func{methods}(C)\\
   x\not\in{\sf bv}(e)\quad
   \Gamma, y'\colon s \vdash e [\thisk\mapsto x][y\mapsto y'] 
   \ret E_2\quad
   y' ~\text{fresh}
  }{
  \Gamma\vdash x.m(sv)\ret 
  E_1 \circ \ottkw{Evar} (\ottkw{VbindC}(y', c[\thisk\mapsto
  x][y\mapsto y'])) :: E_2
  }
  \and
  \inference[\rtit{eUpdate}]{
    \Gamma\vdash sv \ret E_1
  }{
  \Gamma\vdash x.f = sv\ret 
  E_1 \circ [\ottkw{Evar}(\ottkw{Vfield}(x,f))]
  }
  \and
  \inference[\rtit{eCast}]{
    \Gamma\vdash e\colon \ret E
  }{
    \Gamma\vdash (t)e \ret E
  }
  \and
  \inference[\rtit{eLet}]{
    \Gamma\vdash e_1 \ret E_1\qquad  
    \Gamma\vdash e_1 \colon t \qquad
    \Gamma, x' \colon t \vdash e_2[x \mapsto x'] \ret E_2  
    \qquad
    x' \text{ fresh}
  }{
  \Gamma\vdash  \kw{let} \ x\ \c{=} \ e_1\ \kw{in} \ e_2
  \ret 
  E_1 \circ \ottkw{Evar}( \ottkw{VbindL}(x') :: E_2
  }  
  \and
  \inference [\rtit{eExpr}]{
    e = x\text{ or }
    e = v\text{ or }
    e = x.f
    }{
    \Gamma\vdash e \ret [\ottkw{Eret}(e)]
     }
    \and
  \inference[\rtit{eMethod}]{
   \Md = m(x : t_1) : t_2~ [\tilde c]\{e\}\qquad
   \Gamma,x\colon t_1\vdash e \ret E
     }{
     \Gamma \vdash \Md \ret (\tilde c, E)
  }
  
  \end{mathpar}
  \end{stable}
  
The core of the source-level analysis of \model programs is the effects of methods.
%
%
An effect list provides a representation of a program that is executed in a memory-based model. We refer to it as symbolic to distinguish from the actual semantics of \oolong. As in the standard semantics, there is an operation to change the fields of objects, operations to introduce bindings, etc.
The aim is to track the side effects of a method executed abstractly 
in a state where the heap contains symbolic instances of
\thisk and of the formal parameter, dubbed \otherk,
such that  their integer fields are  
aliased by  uninstantiated variables.
%
Interestingly, effects have a \emph{flat} structure, that is, 
lists of effects are symbolically executed in a non-inductive
deterministic framework that relies on
a single memory cell for the return value, thus
providing the ground for efficient implementations.

The syntax of effects and their inference rules are in 
Table~\ref{tab:effects-inference};
the top-level rule for methods, noted $\Gamma\vdash\Md\ret \eff$, 
assumes that $\Md$ belongs to a type-checked program~$P$~\cite{oolong-ACR2019}:
in such case, we say that the method $\Md$ has effect \eff.
In the rules for expressions, where we use $\circ$ for list concatenation,
we consider four {side effects}, or \emph{effect expressions}:
\begin{itemize}
\item $\kw{Evar}$, the effect of a let-expression, of a method call, or of an update;
\item $\kw{Eop}$, the effect of an integer operation;
\item $\kw{Enew}$, the effect of a new;
\item $\kw{Eret}$, the return-only effect of the remaining expressions.
\end{itemize}

 The parameter of $\kw{Evar}$ is an \emph{effect variable}:
the effect variables for let-expressions and
calls receive as parameter a variable~$x$ that is freshly
added by the inference system in order to substitute the occurrences
of the let-variable and of the method's formal parameter, respectively
(cf.~\rtit{eLet} and \rtit{eCall}).
Given the class of the instance, 
in \rtit{eCall} the effect of the call is calculated by 
substituting both the occurrence of $\thisk$ in the body with the caller~$x$,
where we assume that $x$ is not bound in the body (eventually by alpha-renaming),
and the occurrence of the formal parameter with the fresh variable~$y$.
Notably, the effect variable $\kw{VbindC}(y,\tilde c)$ of a method call 
also carries the method's preconditions $\tilde c$ opportunely instantiated
with the actual caller and the fresh variable instantiating the formal parameter:
this allows to track the chain of preconditions originating
from nested method calls, and the actual parameters of the preconditions.
The effect variable for an update provides the variable and the field (cf.~\rtit{eUpdate});
the symbolic value $sv$ to be assigned to the field is the result of the symbolic evaluation of the 
last entry on the left in the 
returned effect list, that is the tail of $E_1$.
For instance, when $sv=sv_1 ~\ottnt{Op}~ sv_2$, 
this would be the tail 
$[\ottkw{Eop}(Op,sv_2)]$,
symbolically evaluated after the evaluation of the effect of $sv_1$ 
(cf.~\rtit{eOp}).
This is the general mechanism to pass values among effects: 
each effect~$b$ in $E\circ b::E'$ ``receives'' the result of the evaluation of the tail of $E$.

\paragraph{Symbolic Semantics}
Recall%
\footnote{Definition in the paragraph on the Dynamic Semantics in \S~\ref{sec:model}.}
that a configuration is a pair with a state and an expression, being the state a fourtuple (heap, stack, a field to monitor method calls' pre-conditions and another to record access to strong fields). Symbolic configurations $e\Cfg$ are composed by: a \emph{symbolic heap}, containing objects that have symbolic values in their field map; a \emph{symbolic stack}, mapping variables to symbolic values; a \emph{field} to monitor the access to strong fields (as in standard configurations); an \emph{effect}; and a \emph{symbolic value}, also called \textit{buffer}, returned by the last effect expression evaluated in the effect list.

Intuitively, the semantics of an effect $\eff$ is based 
on \emph{evaluating it} in a symbolic configuration, a relation noted $e\Cfg \rightarrow e\Cfg'$.
The interested reader will find the evaluation rules in~\S~\ref{sec:effect-rules}.
By executing the effect of a method definition in a symbolic state and consuming all entries of
the effect list, we obtain a state's heap containing the manipulations done by the method's body to the objects in the  heap (which may grow due to \texttt{new} expressions, cf.~ \rtit{eNew}).
Moreover, in order to describe the behaviour of all guarded calls of the method, we require that
the final preconditions are \textit{SAT}.

For a symbolic configuration $e\Cfg$, let $\func{update}_{\kw s}(e\Cfg)= e\Cfg'$ whenever
$e\Cfg\rightarrow^* e\Cfg'$, and $e\Cfg'$ contains the effect $(\tilde c', [])$, and $\tilde c'$ is SAT.
Given the invariants $\tilde c$ and a symbolic state $e\Sigma$,
let the ($e\Sigma$-)closure of $\tilde c= c_1,\dots,c_n$, noted $\tilde c e\Sigma$, be the invariants 
$c_1 e\Sigma,\cdots,c_n e\Sigma$, where each $c_i e\Sigma$ is obtained by substituting the fields in $c_i$ with the values in $e\Sigma$ (cf.~\S~\ref{sec:conflicts}),
$1\leq i\leq n$.
 %
%

\begin{example}\label{ex:effect-execution}
  To illustrate the symbolic evaluation of effects, we symbolically execute method \texttt{init} in Figure~\ref{lst:bank}.

  Let $a,b$ and $m$ abbreviate \texttt{amount}, \texttt{balance} and \texttt{min\_cash}, respectively.
  The effect of the body is $\eff=((a \geq 0,a\leq \thisk.m), E)$, where
  $E=\ottkw{Eret}(a):: \ottkw{Evar}( \ottkw{Vfield}(\thisk,b)]$.

  Consider the heap $[\thisk\mapsto (A, [b\mapsto z][ m\mapsto y])]$.
  To complete the configuration, noted $e\Cfg$, 
  we add an empty stack, empty monitored strong fields, the effect obtained by closing the precondition,
  and the buffer $\nullk$; the closure is $(a\geq 0, a\leq y)$.

  Given a variable $x$, we say that the reduction system \textit{locates} $x$ whenever it finds $x$ in the domain of the stack, or in the domain of the heap.

  The first reduction step  takes the head of the effect list, produces a configuration containing  the precondition and the effect's tail, fails to locate $a$, and in turn sets the buffer to $a$ (where if e.g. we had $eV(a)=sv$, the buffer would have been set to $sv$).

  The second  step produces the precondition and an empty effect list, locate $\thisk$, 
  produces the heap $[\thisk\mapsto (A, [b\mapsto a][ m\mapsto y])]$, and sets the buffer to $\nullk$, thus producing a configuration $e\Cfg'$.

  Now the computation ends, as the effect list is empty. Note that since the reached precondition $(a\geq 0, a\leq y)$ is {SAT}, the  heap of the redex contains the legit modifications done by all guarded calls of \texttt{init}: $\func{update}_{\kw s}(e\Cfg)=e\Cfg'$.
  This would not have been the case if, for instance, the body of \texttt{init} 
  was $\thisk.\texttt{deposit}(-10)$, which would have spawn 
  $(a\geq 0, a\leq y, -10 \geq 0)$ as final precondition.
\end{example}  

\subsection*{Effect-based anticipation analysis}\label{sec:symb-semantics}
The static analysis presented in this section relies on the symbolic representation
of objects: given a class, we generate an instance by generalizing all integer fields 
with symbolic variables, and by instantiating all object fields with 
generated instances of their class.
Given a set of classes, we generate a symbolic state, noted $e\Sigma$, 
by generating a symbolic heap containing symbolic objects,
by adding an empty stack,
and by adding empty monitored strong fields.
In order to generate the heap for the analysis, 
we thus need to determine the set of classes under consideration,
and the \textit{aliases} to objects.
Given two methods definitions $\Md_1,\Md_2$ whose 
method calls are executed in sequence, 
the classes involved are those of $\thisk_1$ and $\thisk_2$, and those of 
$\otherk_1$ and $\otherk_2$ iff each $\otherk_i$ is an object, $i=1,2$;  
each class could recursively spawn further instances by having object fields.
In order to generate the aliases, 
we need to control interference among the two calls of the methods.
To this aim,  we analyze all possible equalities  among 
the locations of the objects instantiating 
$\thisk_1$, $\thisk_2$, $\otherk_1$, and $\otherk_2$, noted $x_1\doteq x_2$.
In general, there are fifteen possible equalities cases,
e.g. 
1) $\thisk_1\doteq\thisk_2$ and $\thisk_2 \doteq \otherk_1$ and
$\otherk_1 \doteq\otherk_2$, or
6) $\thisk_1 \doteq\thisk_2$ and $\otherk_1 \doteq\otherk_2$, or
12) $\thisk_1 \doteq\otherk_2$, and so on.
In particular, e.g. when the type of both the first and the second method's parameter is not a class,
we have at least two cases: $\thisk_1 \doteq\thisk_2$ or $\thisk_1\not \doteq\thisk_2$.
We indicate the equality case under analysis with $\hcase$, 
and write $\kw{gen}(\hcase,\Md_1,\Md_2)$ to indicate the symbolic state 
induced by the generated symbolic heap containing the entries projected by $\hcase$:
e.g., in case 1)  above the heap must contain an object entry aliased by 
$\thisk_i, \otherk_i$, $i=1,2$;
in case 6) it must contain an entry aliased by 
$\thisk_i$, $i=1,2$ and another different entry aliased by 
$\otherk_i$, $i=1,2$;
in case 12) it must contain an entry aliased by
$\thisk_1,\otherk_2$, a different entry for $\thisk_2$, and, 
if $\otherk_1$ is an object,
a different entry for $\otherk_1$.

\paragraph*{Symbolic Method Sequence.}
Consider:
(i) an equality case $\hcase$;
(ii) method definitions $\Md_1$ and $\Md_2$ such that, for each $i \in \{1,2\}$, $\thisk_i\colon C_i\vdash\Md_i\ret (\tilde c_i, E_i)$;
(iii) $e\Sigma = \kw{gen}(\hcase,\Md_1,\Md_2)$;
(iv)) $e\Cfg =\langle e\Sigma, (\tilde c_1 e\Sigma, E_1),\nullk\rangle$.

Assume $\func{update}_{\kw s}(e\Cfg)= e\Cfg'$, where $e\Cfg'$ contains state~$e\Sigma'$
and effect $(\tilde c', [])$.

Furthermore, consider the configuration $e\Cfg_2$ obtained from $e\Cfg'$ by:
(i) substituting the stack with an empty stack, the effect with $(\tilde c_2 e\Sigma', E_2)$, and the buffer with $\nullk$;
and (ii) by adding a separator to the monitored strong fields.

Finally, assume that $\func{update}_{\kw s}(e\Cfg_2)= e\Cfg''$
and that $e\Cfg''$ contains heap $eH$, monitored strong fields $\sff$, and effect $(\tilde c'', [])$.

We can now define the notion \emph{symbolic method sequence}: 
$$\Md_1 ; \Md_2 \lts{\hcase} eH,\sff,(\tilde c',\tilde c'')$$

\paragraph{Symbolic commutativity and local permissibility}
The static  notion of commutative method sequence is parametrized by 
a set of equations over symbolic values:
the equations are the result of comparing the symbolic heaps obtained by 
symbolically executing  a sequence of two methods, and its inverse.
\emph{Commutativity} says under which conditions the two
methods commute, that is it holds when there exists a set of  
equations $eqs$ that \emph{equate} the symbolic heaps $eH'$ and $eH''$ of the two method sequences, respectively, noted $eqs\Vvdash eH_1\sim eH_2$, and when the state invariant and 
the invariants of the two sequences are satisfied.

The \emph{state invariant} $\func{inv}(e\Sigma) = \tilde c e\Sigma$, 
is the set of invariants $\tilde c$ built upon the fields' invariants 
of the objects' classes contained in the heap $eH$ of $e\Sigma$, 
and closed by the actual values of the fields in $eH$.

The \emph{weak state invariant}, noted $\func{weak\_inv}(e\Sigma)$, 
is the subset of the invariants of $\func{inv}(e\Sigma)$ s.t.
their fields are marked as weak in their classes. 

Let $\func{fv}(\tilde c)$ denote the free variables of a set of invariants $\tilde c$.

\paragraph*{Two method definitions $\Md_1$ and $\Md_2$ commute}\label{def:comm-methods}
w.r.t. equality case $\hcase$, equations $eqs$, and
invariants $\tilde c$, noted 
$$eqs;\tilde c\models_\hcase \Md_1 \rightleftarrows\Md_2$$
whenever
\begin{itemize}
  \item $\Md_1 ; \Md_2\lts{\hcase}eH',\sff',\tilde {c'}$;
  \item $\Md_2 ; \Md_1 \lts{\hcase} eH'',\sff'',\tilde{c''}$;
  \item  $eqs\Vvdash eH'\sim eH''$; and
  \item $\tilde c = \func{inv}(\kw{gen}(\hcase,\Md_1,\Md_2)), 
  \tilde {c'}, \tilde {c''}$.
\end{itemize}

The static analysis of local permissibility provides the post-state
invariant
\footnote{State-invariant is a notion presented in the beginning of \S~\ref{sec:conflicts}}
that must hold after the symbolic
execution of the method when the pre-state invariant and the method's preconditions hold:
that is, the analysis returns a proposition (e.g. a term of type \texttt{Prop} in Coq) 
that must be SAT in order to establish local permissibility.

\paragraph*{Static Analysis of Locally Permissible Methods.}\label{def:lp-method}
Let $\Md$ be a method s.t. $\thisk\colon C\vdash\Md\ret (\tilde c, E)$,
and let $e\Cfg = \langle e\Sigma,(\tilde c,E),\nullk\rangle$,
where $e\Sigma$ contains a symbolic heap with an entry $[\thisk\mapsto (C,eF)]$, for some symbolic field map $eF$.
Let $\func{update}_{\kw s}(e\Cfg) = e\Cfg'$ with $e\Cfg'$ containing the state $e\Sigma'$ and 
the effect $(\tilde c',[])$.

We say that:
\begin{itemize}
  \item $\Md$ is \emph{permissible} in $e\Sigma$ up-to $\phi$, 
  noted $\textbf{sP}(e\Sigma,\Md)=\phi$,
  
  whenever $\phi = 
  \kw{inv}(e\Sigma) \land {\tilde c'}\Rightarrow \kw{inv}(e\Sigma')$;
  \item $\Md$ is \emph{not-permissible} in $e\Sigma$ up-to $\phi$, noted $\textbf{sNP}(e\Sigma,\Md)=\phi$, 
  
  whenever $\phi=
  \kw{inv}(e\Sigma)\land{\tilde c'}\Rightarrow\kw{not}(\kw{inv}(e\Sigma'))$.
  \item $\Md$ is \emph{locally permissible} in $e\Sigma$ up-to $\phi$,
  noted
  
  $\textbf{sLP}(e\Sigma,\Md)=\phi$,

  whenever $\phi = \forall x\in\kw{fv}(\kw{inv}(e\Sigma),{\tilde c'})\, . \,
  \textbf{sP}(e\Sigma,\Md)$.
\end{itemize}

\begin{table}
  \caption{Method Anticipation Algorithm}
  \label{tab:static-anticipation-algorithm}
  \ocamlexternal{alg-static.ml}
  \end{table}

\paragraph{The Method Anticipation algorithm}
(in Table~\ref{tab:static-anticipation-algorithm}), to decide when two method calls can be swapped, resembles that in Table~\ref{tab:anticipation-algorithm} but:
(i) is source-level, that is, it analyzes two method definitions without relying on a state;
(ii) returns a list of propositions whose conjunction must be satisfied at runtime.
%

The algorithm works like this:
\begin{enumerate}
  \item after building the initial symbolic state (l2), generates the conditions for commutativity (l3) and include them in the result (l6);
  \item then, it adds to the result the \textbf{sLP} proposition for $\Md_2$ in the initial state (l6), and the disjunction of the \textbf{sLP} proposition for $\Md_1$ in the post state (l7) with a weak fields-quantified conjunction of the implications preserving permissibility in the pre and post states (l8--l10);
  \item finally, checks that $\Md_1$ does not modify the strong fields read by $\Md_2$:
    $\textbf{SFNI}_\hcase (\Md_1,\Md_2)$ holds whenever

    $\Md_1 ;$ $ \Md_2 \Lts{\hcase} eH,(\sff',\kw{sep},\sff''),(\tilde c',\tilde c'')$
    and $(\iota,f,\kw{r})\in\sff''$ imply $(\iota,f,\kw{w})\not\in\sff'$,

    where $\kw r$ and $\kw w$ indicate a read and write access, respectively.
\end{enumerate}

$\textbf{SFNI}_\hcase (\Md_1,\Md_2)$ captures that fact methods  $\Md_1$ and $\Md_2$
do not interfere.

\begin{example}
  Consider the sequence of methods \texttt{getBalance}
  and \texttt{accrueInterest} in Listing~\ref{lst:bank}
  (abbreviated to \texttt g and \texttt a, respectively).
  We want to find the conditions under which a call of \texttt a can be anticipated
  w.r.t.  a call of \texttt g in a (runtime) state $\Sigma$ s.t.
  $\textbf I(\Sigma)$.
 
  Let $z$ and $y$ be the unistantiated variables of the field \texttt{balance} and \texttt{min$\_$cash}, respectively, 
  and let $i$ be a shorthand for the formal parameter of \texttt a.
 
  The algorithm in Table~\ref{tab:static-anticipation-algorithm}
  calculates the following values to populate the following list of propositions:
  
  $eqs = (z+z/100*i=z+z/100*i, y=y)$; 
  
  $cc = (z \geq 0, y = 50)$;
  
  $\textbf{sLP}(e\Sigma, \texttt a) ==$
  
  $\quad  \forall z,y,i\, .\,(z \geq 0 \land y = 50)
  \Rightarrow (z+z/100*i \geq 0 \land y = 50)$;
  
  $\textbf{sLP}(e\Sigma'', \texttt g) =$
  
  $\quad \forall z,y,i\, .\,
   (z+z/100*i \geq 0 \land y = 50)\Rightarrow (z+z/100*i  \geq 0 \land y = 50)$;
   
  $\textbf{SFNI}(\hcase,\texttt w,\texttt d)$
  (we omit the remaining values).

  For any state $\Sigma$ preserving its invariant, all invariants $c\in\eqs\Sigma$ are SAT, as well as the invariants in $cc\Sigma$.

  Next, we need to ensure that $\textbf{sLP}(e\Sigma, \texttt a)$ is SAT:
  this holds whenever $i\geq -100$. 
  Since $\textbf{sLP}(e\Sigma'', \texttt g)$ is SAT, the disjunction is SAT as well;
  the last entry of the list is trivially SAT.
  Therefore, \texttt{accrueInterest} can be anticipated w.r.t. \texttt{getBalance}
  when the interest is greater or equal to $-100$.
\end{example}

\section{A Java Implementation}~\label{sec-java-impl}
To assess the use of \model in real-life examples, 
we implemented a prototype that analyses Java Bytecode.
It is a adaptation/generalization of the analysis presented in \S~\ref{sec:re-static-commutativity} to cope with Java Bytecode instruction set (including conditions and jumps) and the virtual machine's stack-based addressing mode.
Weak fields an d field invariants are expressed through Java annotations. Preconditions have been passed over by if instructions.

To generate and solve the equation system, we make use of CVC4~\cite{DBLP:conf/cav/BarrettCDHJKRT11} by the means of the Servois~\cite{DBLP:journals/jar/BansalKT20} commutativity condition generator.
%
%
The analysis generates an anticipation table (similar to Table~\ref{tab:results}) with the calls that may cause conflicts and, for each of these, the conditions on which a call may anticipate another.
The table is implemented as a map from method signatures ($m_2$) onto another map, 
whose keys are the signatures of the methods ($m_1$) with which $m_2$ conflicts
and the values are the conditions (as Java lambda functions) for $m_2$ to anticipate $m_1$.
These lambda functions have, as parameters, the list of arguments of each method.

We assessed how much time it takes to check anticipation. 
Although the implementation does not restrict the number of arguments of methods and the complexity of a query depends, naturally, on the complexity of the condition to check, our experimental results show querying the table takes less than $0.001~ms$, several orders of magnitude lower than performing global synchronization. We conclude that it is feasible to use our approach at runtime.

\section{Use-Cases}\label{sec:evaluation}
\begin{table*}[t]
  \caption{Statistics on the use-cases. 
   \antc{m_2}{m_1}{c} indicates that, given a sequence $m_1;m_2$,  method $m_2$ can anticipate $m_1$ when $c$ holds, thus giving rise to the sequence $m_2;m_1$.}
\label{tab:results}
\begin{small}
\begin{tabular}{l||r|r|r|r
    ||r|r|r}
\multirow{2}{*}{\textbf{Use-case}} & 
\# \weakk / & 
\multirow{2}{*}{\#   \textbf{invs}}   & 
\multirow{2}{*}{\#  \textbf{mtds}} & 
\# \multirow{2}{*}{\textbf{non-LP}}   & 
\multirow{2}{*}{\# \textbf{pairs}} &
\multirow{2}{*}{\# \textbf{conflicts}} &
\multirow{2}{*}{\# \textbf{anticipations}} 
\\
&  \# \textbf{strong} &   &  &     & &
\\        
\hline
  Account    & 1/1 & 2 & 6 & 3 & 21 & 5 
                            & \antc{\texttt a}{\texttt g}{i\geq-100},
                              \antc{\texttt a}{\{\texttt d, \texttt t, \texttt w, \texttt a\}}{\thisk_1\not\doteq\thisk_2\land i_2\geq -100},
  \\
    &   &  &  &     & &&                                      
                                     \ant{\texttt g}{\texttt *},
                                 \ant{\texttt d}{\{\texttt d, \texttt g\}},
                                     
                     \antc{\texttt d}{\{\texttt t, \texttt w, \texttt a\}}{\thisk_1\not\doteq\thisk_2}
\\
\hline
  Auction & 1/1 & 2 & 4 & 0  & 9 & 3 
                              & 
                    \ant{\texttt b}{\texttt {cb}},
                    \ant{ \texttt{cb}}{\texttt {*}},
                    \ant{ \texttt{c}}{\texttt{c}},
   \ant{\texttt b}{\texttt {b}}
\\
\hline
  Counter & 1/0 & 0 & 3 & 0 & 6 & 0  
                                           &
                            \ant{\texttt *}{\texttt *}
\\
\hline
  Register & 3/0 & 0 & 2 & 0 & 3 & 0 
  &
                  \ant{\texttt{g}}{\texttt{*}},
                \ant{\texttt{*}}{\texttt{g}},
     \ant{\texttt{s}}{\texttt{s}},
    \\
  \hline
\end{tabular}
\end{small}
\end{table*}

Table \ref{tab:results} presents statistics of applying\ model to common use-cases. For each,
the number of weakly consistent fields, of field invariants, of methods, as well as the result of the \textbf{\model analysis}: number of non-{LP} methods, of pairs of methods analyzed, of conflicts (non-commutative methods) detected, and possible anticipations.
%

\paragraph{Account}
Consider the example presented in Listing~\ref{lst:bank}.
Method \texttt{init} (\texttt{i}, for short) has been replaced by a Java  constructor. Constructors are always LP and do not generate conflicts against methods invoked on the object itself. 
The results of the static analysis are:
\begin{description}
\item[non-LP methods:]
\texttt{transfer} (\texttt{t}, for short), \texttt{withdraw} (\texttt{w}), and \linebreak\texttt{accrueInterest} (\texttt{a}).
\item[non-commutative methods:] when applied on the same account,
 \texttt{deposit} (\texttt{d}), \texttt{withdraw}, \texttt{transfer} and \texttt{accrueInterest} itself with \texttt{accrueInterest}.
\item[anticipatable methods:] \texttt{getBalance} (\texttt{g}) can anticipate all method calls;
\texttt{deposit} can anticipate calls of LP methods;
 to anticipate \texttt{deposit} w.r.t. a non-LP method call, the two accounts must differ;
 to anticipate \texttt{accrueInterest}, the interest must be greater or equal~$-100$;
 and, when the call preceding \texttt{accrueInterest} is not a \texttt{getBalance}, the two accounts must differ.
\end{description}

\paragraph{Auction} An Ebay-like auction 
where bid operations operate over a weak state, while the closing (and deciding the winner) operates over a strong state. Once the auction is closed, bid operations have no impact. The operations are: \texttt{bid} (\texttt{b}) to place a new bid (that must be higher than the current),
\texttt{currentBid} (\texttt{cb}), 
that returns the value of current highest bid,  
\texttt{close} (\texttt{c}), that closes the auction, 
and \texttt{winner} (\texttt{w}) that returns the identifier of the winning bidder. \texttt{close} and \texttt{bid}  only have side effects if the \textit{auction is open}. 
An auction's state comprises  two replicated fields of type \texttt{Bid} (with the bidder and the bid):
the weakly consistent \texttt{current\_bid} and 
the \textit{strong} \texttt{winner} indicating the auction's winner (with a special value to indicate that the auction is still open).
%
%
The results of the static analysis are:
\begin{description}
\item[non-LP methods:] none.
\item[non-commutative methods:] \texttt{bid} with \texttt{close}, when applied upon the same auction.
\item[anticipatable methods:]
\texttt{currentBid} and \texttt{bid} can anticipate each other, \texttt{close} can anticipate itself,
\texttt{bid} can anticipate itself because 
two consecutive bids  produce the same outcome: the current bid becomes the highest of the two.
All methods can anticipate any other method of different auctions.
\end{description}

\paragraph{Counter} An eventually consistent Positive-Negative (PN) counter 
(like a PN-counter CRDT~\cite{crdts-SSS2011}), with operations to \texttt{increment} (\texttt{i}), \texttt{decrement} (\texttt{d}) and \texttt{read} (\texttt{r}).
%
The results of the static analysis are:
\begin{inparadesc}
\item[non-LP methods:] none.
\item[non-commutative methods:] none.
\item[anticipatable methods:] all (as expected in a CRDT-like datatype). 
\end{inparadesc}

\paragraph{Register}
A Last-Writer-Wins Register with two base operations: \texttt{set(value, sequenceNumber, replicaId)} 
and \texttt{get}(), and a second set operation
\texttt{set(value)} that calls

{\hspace{20pt} \texttt{set(value, seqNumber++, localReplicaId)}.}

\noindent Each replica features three weak replicated
fields: \texttt{value}, with the register's  value, 
\texttt{seqNumber}, with the sequence number 
of the last update to field \texttt{value}, 
and \texttt{replicaId}, with
the identifier of the replica that emitted such update. \texttt{localReplicaId} is an not replicated constant field with the identifer of the local replica.
The results of the static analysis are:
\begin{inparadesc}
\item[non-LP methods:] none.
\item[non-commutative methods:] none.
\item[anticipatable methods:] 
\texttt{get} can always anticipate, as well as be anticipated, 
\texttt{set} can anticipate itself.
 %
\end{inparadesc}

\section{\model and Coordination Synthesis}
\label{sec:cs}

In this section we present how \model
can contribute to the synthesis of the  
coordination level needed to execute a given method call. 


\textit{Strong consistency.}
A method call is a strongly consistent operation requiring distributed coordination
if it either: (i) accesses a strong field,
or (ii) breaks the invariant of at least one weak field 
(it is not LP).
Examples of (i) are calls to method \texttt{close} of \texttt{Auction}.
As expected, no method call that reads or writes the same auction can be anticipated w.r.t. \texttt{close} and vice-versa (see Table~\ref{tab:results}).
The same happens to the calls that fall in category (ii) whenever an invariant may be broken.
Examples are calls
to \texttt{withdraw} of \texttt{Account}.

\textit{Causal consistency.}
Under causal consistency, causally related method calls cannot anticipate one another.
Causal relations and session guarantees are orthogonal to the notion of call anticipation, and should be treated upstream. Given two methods calls $mc_1$ and $mc_2$, if $mc_1$ is under coordination and there are \textbf{no} causal dependencies between both calls,  the table produced by \model can be used to check if $mc_2$ may anticipate $mc_1$.



\textit{Eventual consistency.}
Every LP method call that only accesses weak fields  
and whose side effects do not depend on the effects of 
any other call may execute with eventual consistency.
Trivial examples are 
calls to the methods from a weakly consistent PN-counter that features a weak single field.

\section{Related Work}

In Kraska et al.~\cite{cr-PVLDB2009}, DCCT~\cite{dccp-PLMDC2016} and MixT~\cite{mixt-PLDI2018}, as in \model, consistency guarantees are bound to data.
However, in these proposals, 
the programmer must  reason about the  consistency levels, explicitly binding data 
to consistency categories, such as \textit{serializable} or \textit{eventual}, which contrasts with \model's \texttt{weak}.
%
All of them  preclude the coexistence of operations with different consistency levels over the same data.
Moreover, no reference is made to which operations/transactions may execute concurrently,  \textit{i.e.} to call anticipation.

%
%
%

Red-blue consistency \cite{redblue-OSDI2012} colors operations, red or blue, according to whether 
they require global coordination or not.
Thus, contrarily to \model, consistency guarantees are attached to operations.
Although the work provides a set of rules 	for labeling operations blue, 
it is up to the programmer to  reason about operation commutativity and invariant violation.
%
Gotsman et al.~\cite{se-POPL2016} go a step further, providing a proof rule for 
assessing if a choice of consistency guarantees 
for a set of operations over a replicated state is sufficient to preserve the  data integrity invariant.
Also, in red-blue,  all operations that are not globally commutative must be totally ordered.
Partial Order-Restrictions  consistency (Olisipo)~\cite{fgc-ATC2018}  tackles this issue by 
allowing for the definition of a set of restrictions that must be met in all partial orders of the operations.
%
\model establishes partial order relationships by design, as operations operating over different data objects do not conflict with each other.

Sieve \cite{sieve-ATC2014} 
automates the choice of the red-blue consistency levels for operations over replicated databases. 
Programmers need to annotate the database's schema with the CRDTs~\cite{crdts-SSS2011} 
that encapsulate the merge semantics of a table or of a table field. 
%
%
%
Sieve and Olisipo  use Weakest Pre-condition Calculi to find restrictions for preserving invariants,
and use commutativity analysis to compute which operations may execute concurrently. To the best of our knowledge, as in Hamsaz, the output is true or false, which is a weaker result than our static analysis that is able to output parameterized conditions that may be instantiated at runtime to check for conflicts.

Quelea~\cite{quelea-PLDI2015} and  Indigo~\cite{ec-Eurosys2015}
propose contract-based approaches to specify consistency requirements. 
Given the specifications both strategies identify which operations may, or may not, safely execute concurrently.
In Quelea, the programmer must reason about the effects of each operation/transaction and how these interfere with each other.
%
Failing to identify a conflict will result in a contract that, although correct, does not reflect the necessary constraints.
\model assumes some of this burden, inferring the effects from the annotated data and detecting possible conflicts by conducting a static  analysis.
%
%
In Indigo, the programmer must provide the list of  invariants over the application's state and a list of effects per operation.
The programmer does not have to reason about effect interference, 
but rather about which are the effects produced by each operation.
A single missing or miswritten effect may render the specification  incorrect.
%
\model has thus the advantage of 
automating the inference of the effects 
and performing the subsequent invariant violation and conflict analyses.
On the other hand, Indigo 
allows to specify invariants that relate multiple data items.

\textsc{AutoGr}~\cite{autogr} provides an automated way of geo-replicate non-replica\-ted applications, relying on the framework \textsc{Olisipo} and using the static analysis tool Rigi to automatically identify potentially conflicting operations. \model works directly on the source code and also defines a static analysis, but not only to detect conflicts: it provides the run-time with information of which operations may be anticipated as no coordination is required.

\section{Conclusions}
This paper presents \model, a novel language-based call anticipation static analysis for applications with mixed consistency. The main notions are formally defined in \S~\ref{sec:re-static-commutativity}.
We develop \model over \oolong, an object-oriented core programming calculus, rigorously defined and 
mechanized to ensure type safety. 
The static analysis we devise and present herein infers the effects of the language's methods, and uses such effects to check which methods are locally permissible, which pairs of methods commute and, ultimately, in which conditions method calls may be anticipated.
All these notions were implemented and type-checked in a functional language and are publicly available. 

Our proposal distinguishes itself from the state-of-the-art 
by
enabling the anticipation of method calls: it provides, for each pair of methods, a system of inequalities over objects’ fields and methods’ parameters that must be satisfied at runtime. 
To illustrate the applicability and expressiveness of our approach, \S~\ref{sec:evaluation} presents the output of the analysis for a set of significant examples commonly used in the literature. The results demonstrate the viability of the methodology we devised. Moreover, a Java implementation (presented in \S~\ref{sec-java-impl}) further strengths the usability of the analysis.

Finally, we discuss how \model can be used in the synthesis of  the coordination required 
for each method call, so that the modifications to replicated state comply to 
the consistency annotations provided in the source code (see \S~\ref{sec:cs}). 

In future work, we will mechanise the static analysis soundness, 
and add constructs like conditionals and loops to the 
language.




\bibliographystyle{ACM-Reference-Format}
\bibliography{tech-report} 

\newpage
\appendix
\section{Typing Rules}\label{sec:typing-rules}
 \begin{table}
  \caption{Typing of expressions \framebox{$\Gamma\vdash e \colon t$}}
  \label{tab:ts-expr}

\begin{mathpar}
 \colorbox{lightgray}{
  \inference [\rtit{tLoc}]{
    \vdash \Gamma \qquad
    \Gamma(\iota) = t
}{
\Gamma \vdash \iota\colon  t
}}
\and
\inference [\rtit{tInt}]{
  \vdash \Gamma \qquad n\in\intk
}{
\Gamma \vdash n\colon  \intk
}
\and
\inference [\rtit{tNull}]{
  \vdash \Gamma \qquad
t \ne \intk 
}{
\Gamma \vdash \nullk\colon  t
}
\and
  \inference [\rtit{tVar}]{
    \vdash \Gamma
  }{
  \Gamma,x\colon t \vdash x\colon  t
  }
  \and
 \inference[\rtit{tCast}]{
 \Gamma\vdash e\colon s \qquad  s <: t
}{
 \Gamma\vdash (t)e\colon t
}
\and
\inference [\rtit{tOp}]{
  \Gamma\vdash sv_1\colon \intk \qquad
 \Gamma\vdash sv_2\colon \intk
}{
\Gamma \vdash sv_1 ~\ottnt{Op}~ sv_2 \colon \intk 
}
    \and
  \inference[\rtit{tNew}]{
    \vdash \Gamma\qquad
     \kw{class}\ C\, \kw{implements}\, I \SB \_
    \FB\in P
  }{
  \Gamma\vdash\kw{new} \ C \colon C
  }
\and
\inference[\rtit{tCall}]{
\Gamma\vdash x\colon t
\qquad
\Gamma \vdash sv \colon s
\qquad
m(y : s) : t' \,[\tilde c]\in\func{msigs}(t)
}{
\Gamma\vdash x.m(sv)\colon t' 
}
  \and
  \inference[\rtit{tLet}]{
    \Gamma\vdash e_1 \colon t'\qquad
    \Gamma, x \colon t' \vdash e_2\colon t
  }{
  \Gamma\vdash  \kw{let} \ x\ \c{=} \ e_1\ \kw{in} \ e_2
  \colon t
  }
    \and
  \inference[\rtit{tSelect}] {
    \Gamma\vdash x\colon C  \qquad
   f: t \sim\hspace{-.2em}\weakk~[\tilde c] \in \kw{fields}(C) 
  }{
   \Gamma\vdash x.f\colon t 
  }
\and
\inference[\rtit{tUpdate}]{
\Gamma\vdash x\colon C \qquad
\Gamma\vdash sv\colon t \qquad
f: \sim\hspace{-.2em}\weakk~[\tilde c] \in \kw{fields}(C)
}{
\Gamma\vdash x.f = sv\colon \kw{Unit}
}
\end{mathpar}
\end{table}

   \begin{table}  
    \caption{Typing of programs \framebox{$\vdash P\colon t$}}
    \label{tab:ts-program}
  \begin{mathpar}
  \inference[\rtit{T-wfClass}]{
    \kw{class}\ C \ \kw{implements}\ I \SB \_ \FB \in P 
  }{
   \vdash C 
  }
  \and  
  \inference[\rtit{T-wfInterface}]{
    \kw{interface}\ I \  \SB \_ \FB \in P 
  }{
   \vdash I
  }
  \and  
  \inference[\rtit{T-wfInterfaceE}]{
    \kw{interface}\ I \  \kw{extends}\ I_1,I_2  \in P 
  }{
   \vdash I
  }
  \and  
  \inference[\rtit{T-wfBase}]{
    t\in\{\mathbf{Unit},\intk\} 
  }{
   \vdash t
  }
  \and
  \inference[\rtit{wfFieldInv}]{
    \Gamma \vdash d_1\colon \intk \qquad
    \Gamma \vdash d_2\colon \intk\qquad
    \vdash t
    }{
    \Gamma\vdash d_1 \ \Rel\ d_2
  }
  \and
  \inference[\rtit{wfField}]{
    \thisk\colon C \vdash c_1, \cdots,
    \thisk\colon C \vdash c_n
    }{
    \thisk\colon C\vdash f: t \sim\hspace{-.2em}\weakk~[c_1,\dots,c_n]
  }
\and
\inference[\rtit{wfMethod}]{
  \func{fields}(\tilde c)\subseteq \func{strongFields}(C)\qquad
  \thisk\colon C,x\colon t_1\vdash e
}{
\thisk\colon C\vdash m(x : t_1) : t_2~ [\tilde c]\{e\}
}
  \and
  \inference[\rtit{wfClass}]{
     \forall \Msig\in\kw{msigs}(I).\exists e. \Msig\{e\}\in \Mds
     \\
     \forall \Fd\in\Fds.\thisk\colon C\vdash\Fd 
     \\
     \forall \Md\in\Mds.\thisk\colon C\vdash \Md 
     }{
  \vdash \kw{class}\ C \ \kw{implements}\ I \SB \Fds~\Mds \FB 
  }
  \and
  \inference[\rtit{wfInterface}]{
     \forall m(x : t_1) : t_2$ $[\tilde c]\in\kw{msigs}(I).
     \vdash t_i \ (i=1,2)
     }{
  \vdash \kw{interface}\ $I$ \SB \Msigs \FB 
  }
  \and
  \inference[\rtit{wfInterfaceE}]{
     \vdash I_1 \qquad\vdash I_2
     }{
  \vdash \kw{interface}\ $I$ \ \kw{extends}\ I_1,I_2 
  }
  \and
  \inference[\rtit{wfProgram}]{
  \forall \Id\in\Ids.\vdash \Id \qquad
  \forall \Cd\in\Cds.\vdash \Cd \qquad
  \epsilon\vdash e\colon t
  }
  {\vdash \Ids~\Cds~e \colon t}
  \end{mathpar}
  \end{table}
  
 The typing rules for expressions and programs are in 
 Tables~\ref{tab:ts-expr},~\ref{tab:ts-program}, respectively.
 \begin{table}
    \caption{Typing of configurations \framebox{$\Gamma\vdash \Cfg\colon t$}}
    
  \label{tab:ts-conf}
\begin{mathpar}
    \inference[\rtit{wfCfg}]{
    \Gamma\vdash H\qquad
    \Gamma\vdash V\qquad
    \Gamma\vdash e\colon t\\
    \forall (\iota,f,\_)\in\sff\,.\,  f\in\dom(\snd(H(\iota)))
    \qquad
    \kw{fv}(\tilde c)=\emptyset}{
    \Gamma\vdash \langle H, V, \sff, \tilde c, e\rangle\colon t
    }
    \and
    \inference[\rtit{wfHeap}]{
    \forall \iota\colon C\in\Gamma.
    H(\iota)=(C,F)\land\Gamma;C\vdash F\\
    \forall\iota\in\dom(H).
    \iota\in\dom(\Gamma)\qquad
    \vdash\Gamma    
    }{
    \Gamma\vdash H    
    }
    \and
    \inference[\rtit{wfFields}]{
    \func{fields}(C)=f_1 : t_1 \sim\hspace{-.2em}\weakk~[\tilde c_1],\dots,
    f_n : t_n \sim\hspace{-.2em}\weakk~[\tilde c_n]
    \\
    \Gamma\vdash v_1\colon t_1,\cdots,
    \Gamma\vdash v_n\colon t_n
    }{
    \Gamma;C\vdash f_1\mapsto v_1,\cdots,f_n\mapsto v_n    
    }
\end{mathpar}
\end{table}

 The typing rules for configurations recast those of ~\cite{oolong-ACR2019};
Table~\ref{tab:ts-conf} show a selection of such rules.

\section{Effect Evaluation Rules}\label{sec:effect-rules}
\begin{table*}
    \caption{Symbolic semantics of \model-\oolong.}
    \label{tab:selected-symb-semantics} 
    \hspace{-17em}\textit{Syntax of symbolic executions.}\qquad
      \begin{tabular}{lcll}
      \eCfg{} & $::=$ & $\langle eH, eV, \sff, \eff, eT\rangle$ &\hfill
                                                  \textit{(eConfiguration)}
        \\
           $eH$ & $::=$ & $\epsilon$
                 $~|~$ $eH, \iota \mapsto \textit{eobj}$ &\hfill \textit{(eHeap)} \\
           $eV$ & $::=$ & $\epsilon$ $~|~$ $eV, x \mapsto sv$ &\hfill
                                                                \textit{(eStack)}
        \\
        $eT$ & $::=$ & $sv$ $~|~$ $\kw{EXN}$ 
        &\hfill\textit{(Single eThreads)}
         \\
      $\textit{eobj}$ & $::=$ & $(C, eF)$ 
         &\hfill \textit{(eObject)}\\
           $eF$ & $::=$ & $\epsilon$
                          $~|~$ $e F, f \colon t\ [c]\mapsto sv$\qquad\qquad\qquad  
                          &\hfill \textit{(eField Map)}
      \\                                                                  
      $\rho$  & $::=$ & $\epsilon$   
        $~|~$ $\rho, x \mapsto \iota$         
        &\hfill \textit{(Reserved Locations)} 
      \end{tabular}
         \\
     \hspace{-38em}\textit{Reduction rules}\qquad\framebox{$\eCfg \ltsrho{} \eCfg'$}
    \begin{mathpar}
      \inference[\rtit{symbVal}]{
        h = \kw{Eret}~ v\qquad
        \eff = (\tilde c, h :: E)\qquad
        \eff' =(\tilde c, E)
      }{
        \langle eH, eV, \sff, \eff, sv  \rangle
        \ltsrho{}
        \langle eH, eV, \sff, \eff', v  \rangle
      }
      \and
      \inference[\rtit{symbVarL}]{
        h = \kw{Eret}(x)\quad
        \locate(eV,\rho,x) = \iota\\
        \eff = (\tilde c, h :: E)\quad
        \eff' =(\tilde c, E)
      }{
        \langle eH, eV, \sff, \eff, sv  \rangle
        \ltsrho{}
        \langle eH, eV, \sff, \eff', \iota \rangle
      }
        \and
      \inference[\rtit{symbVar}]{
        h = \kw{Eret}~ x\quad
        eV(x) = sv'\\
        \eff = (\tilde c, h :: E)\quad
        \eff' =(\tilde c, E)
      }{
        \langle eH, eV, \sff, \eff, \eqs, sv  \rangle
        \ltsrho{}
        \langle eH, eV, \sff, \eff', \eqs, sv' \rangle
      }
      \and
      \inference[\rtit{symbVarS}]{
        h = \kw{Eret}(x)\quad
        x\not\in\dom(\rho)\cup\dom(eV)\\
        \eff = (\tilde c, h :: E)\quad
        \eff' =(\tilde c, E)
      }{
        \langle eH, eV, \sff, \eff, sv  \rangle
        \ltsrho{}
        \langle eH, eV, \sff, \eff', x \rangle
      }
    \and
      \inference[\rtit{symbOp}]{
        h = \kw{Eop}~ \ottnt{op}~v_2
        \qquad
        sv = sv_1 ~\ottnt{op}~v_2\\
        \eff = (\tilde c, h :: E)\qquad
        \eff' =(\tilde c, E)
      }{
        \langle eH, eV, \sff, \eff, sv_1  \rangle
        \ltsrho{}
        \langle eH, eV, \sff, \eff', sv \rangle
      }
      \and
      \inference[\rtit{symbOpV}]{
        h = \kw{Eop}~ \ottnt{op}~x_2
        \qquad
        eV(x_2)=sv_2
        \qquad
        \qquad
        sv = sv_1 ~\ottnt{op}~sv_2\\
        \\
        \eff = (\tilde c, h :: E)\qquad
        \eff' =(\tilde c, E)
      }{
        \langle eH, eV, \sff, \eff, sv_1  \rangle
        \ltsrho{}
        \langle eH, eV, \sff, \eff', sv \rangle
      }
      \and
      \inference[\rtit{symbOpS}]{
        h = \kw{Eop}(\ottnt{op},x_2)
        \qquad
        x_2\not\in\dom(eV)\qquad
        sv = sv_1 ~\ottnt{op}~x_2
        \\
        \eff = (\tilde c, h :: E)\qquad
        \eff' =(\tilde c, E)
      }{
        \langle eH, eV, \sff, \eff, sv_1  \rangle
        \ltsrho{}
        \langle eH, eV, \sff, \eff', sv \rangle
      }
      \and
      \inference[\rtit{symbSelect}]{
        h = \kw{Eret}~ x ~f \qquad
        \locate(eV, \rho, x) = \iota
        \\
        eH (\iota)= (C, eF)\qquad
        eF (f) = sv'\qquad
        \sff' =\sff,(\iota, f) 
        \\
        \eff = (\tilde c, h :: E)\qquad
        \eff' =(\tilde c, E)
      }{
        \langle eH, eV, \sff, \eff, sv  \rangle
        \ltsrho{}
        \langle eH, eV, \sff', \eff', sv' \rangle
      }
      \and
      \inference[\rtit{symbUpdate}]{
        h = \kw{Evar}(\kw{Vfield}(x,f)) \qquad
        \locate(eV,\rho,x) = \iota\qquad
        eH (\iota) = (C, eF) \\
        eH' = eH [\iota\mapsto (C, eF [f\mapsto sv])]
        \qquad
        \sff' = f\in{\sf wfds}(C)\,?\,  \sff : \sff,(\iota, f)
        \\
        \eff = (\tilde c, h :: E)\qquad
        \eff' =(\tilde c, E)
      }{
        \langle eH, eV, \sff, \eff, sv  \rangle
        \ltsrho{}
        \langle eH', eV, \sff', \eff', \nullk \rangle
      }
      \and
      \inference[\rtit{symbNew}]{
        h = \kw{Enew}~ C ~\Fds\qquad \iota \text{ fresh}
        \\
        eH' = eH [\iota\mapsto(C, \ottnt{create}~ \Fds)
        \\
        \eff = (\tilde c, h :: E)]\qquad
        \eff' =(\tilde c, E)
      }{
        \langle eH, eV, \sff, \eff, sv  \rangle
        \ltsrho{}
        \langle eH', eV, \sff, \eff', \iota \rangle
      }
      \and
      \inference[\rtit{symbBindL}]{
        h = \kw{Evar}(\kw{VbindL}~x)  \qquad
        eV' = eV [x \mapsto sv]\qquad
        \\
        \eff = (\tilde c, h :: E)\qquad
        \eff' = (\tilde c, E)
      }{
        \langle eH, eV, \sff, \eff, sv  \rangle
        \ltsrho{}
        \langle eH, eV', \sff, \eff', \nullk \rangle
      }
      \and
      \inference[\rtit{symbBindC}]{
        h = \kw{Evar}(\kw{VbindC}(y, c))\qquad
        \qquad
        eV' = eV [y\mapsto sv]
     \\
    c_1 = \kw{closure}_{eH,eV'}(c)
    \qquad 
        \eff = (\tilde c, h :: E)\qquad
        \eff' = ((\tilde c,c_1), E)
      }{
        \langle eH, eV, \sff, \eff, sv  \rangle
        \ltsrho{}
        \langle eH, eV', \sff, \eff', \nullk \rangle
      }
    \end{mathpar}
    \end{table*}
The rules to evaluate the effect of methods presented 
in~\S~\ref{sec:static-commutativity}
are in Table~\ref{tab:selected-symb-semantics}.

\end{document}